\documentclass[english]{IEEEtran}
\usepackage[T1]{fontenc}
\usepackage[latin9]{inputenc}
\usepackage{amsmath}
\usepackage{amssymb}
\usepackage{graphicx}
\usepackage{esint}

\makeatletter

\providecommand{\tabularnewline}{\\}

\makeatother

\usepackage{babel}
\begin{document}

\title{Natural Steganography:\\ cover-source switching for better steganography }

\author{Patrick Bas}
\maketitle
\begin{abstract}
This paper proposes a new steganographic scheme relying on the principle
of ``cover-source switching'', the key idea being that the embedding
should switch from one cover-source to another. The proposed implementation,
called Natural Steganography, considers the sensor noise naturally
present in the raw images and uses the principle that, by the addition
of a specific noise the steganographic embedding tries to mimic a
change of ISO sensitivity. The embedding methodology consists in 1)
perturbing the image in the raw domain, 2) modeling the perturbation
in the processed domain, 3) embedding the payload in the processed
domain. We show that this methodology is easily tractable whenever
the processes are known and enables to embed large and undetectable
payloads. We also show that already used heuristics such as synchronization
of embedding changes or detectability after rescaling can be respectively
explained by operations such as color demosaicing and down-scaling
kernels. 
\end{abstract}

\section{Introduction}

Image steganography consists in embedding a undetectable message into
a cover image to generate a stego image, the application being the
transmission of sensitive information. As Cachin proposed in~\cite{Cachin:1998:ITM},
one theoretical approach proposed for steganography is to minimize
a statistical distortion, and the author proposes to use the Kullback-Leibler
divergence. It is interesting to note that this line of research has
been rarely used as a steganographic guideline with few notable exceptions
such as model-based steganography~\cite{Sallee03} which mimics the
Laplacian distributions of DCT coefficients during the embedding,
HUGO~\cite{ih10Pevny} whose model-correction mode tries to minimize
the difference between the model of the cover image and the stego
image, and more recently the mi-pod steganographic scheme~\cite{VS2016TIFS}
which minimizes a statistical distortion (a deflexion coefficient)
between normal distributions of cover and stego contents. 

Currently the large majority of steganographic algorithms are based
on the use of a distortion (also called a cost) which is computed
for each pixel, and which is combined with a coding scheme that minimizes
the global distortion while embedding a given payload. Classical distortions
functions such as the ones proposed by S-UNIWARD~\cite{holub2014universal}
or by HILL~\cite{li2014new} try to infer the detectability of each
pixel by assigning small costs to pixels that are difficult to predict
(usually textural parts of the image) and by assigning large costs
to pixels that are easy to predict (belonging to homogeneous areas
and to some extend to edges). Note that a recent trend of research~\cite{denemark2014selection,li2015strategy}
proposes to correlate embedding changes on neighboring pixels by adjusting
the cost w.r.t the history of the embeddings performed on disjoint
lattices in order to decrease the detectability on greyscale images
or on color images~\cite{tang2016clustering}. 

Once the distortion is computed, a steganographic scheme can either
simulate the embedding by sampling according to the modifications
probabilities $\pi_{k},$ $k\in[1,\dots,Q]$ for a Q-arry embedding,
or can directly embed the message using Syndrome Trellis Codes (STCs)~\cite{filler2011minimizing}
or multilayer STCs~\cite{filler2011minimizing,IHACM16-multilayerSTC}.
The size of the embedding payload $N$ is computed as $N=\sum\pi_{k}\log\pi_{k}$
for each pixel of the image, and in practice the STCs succeed to reach
90\% to 95\% of the capacity~\cite{filler2011minimizing} and consequently
are close to optimal.

Another ingredient to tend to undetectable steganography is to use
the information contained in a ``pre-cover'', i.e. the high resolution
image that is used to generate the cover at a lower resolution, in
order to weight the cost w.r.t the rounding error. For quantization
or interpolation operations, a pixel of the pre-cover at equal distance
between two quantization cells will have a lower cost than a pre-cover
pixel very close to one given quantization cell. This strategy has
been used in Perturbed-Quantization~\cite{fridrich2004perturbed}
but also adapted in more recent schemes using side information~\cite{denemark-wifs2016}. 

The proposed paper uses similar ingredients shared by modern steganographic
methods, namely model-based steganography, Q-arry embedding and the
associated modification probabilities $\pi_{k}$, and side-information.
The main originality of this paper relies on the possible definitions
of cover sources and the use of \emph{cover-source switching} to generate
stego contents whose statistical distributions are very close to cover
contents.

\subsection{What's a source?}

\begin{figure*}[t]
\begin{centering}
\includegraphics[width=0.8\paperwidth]{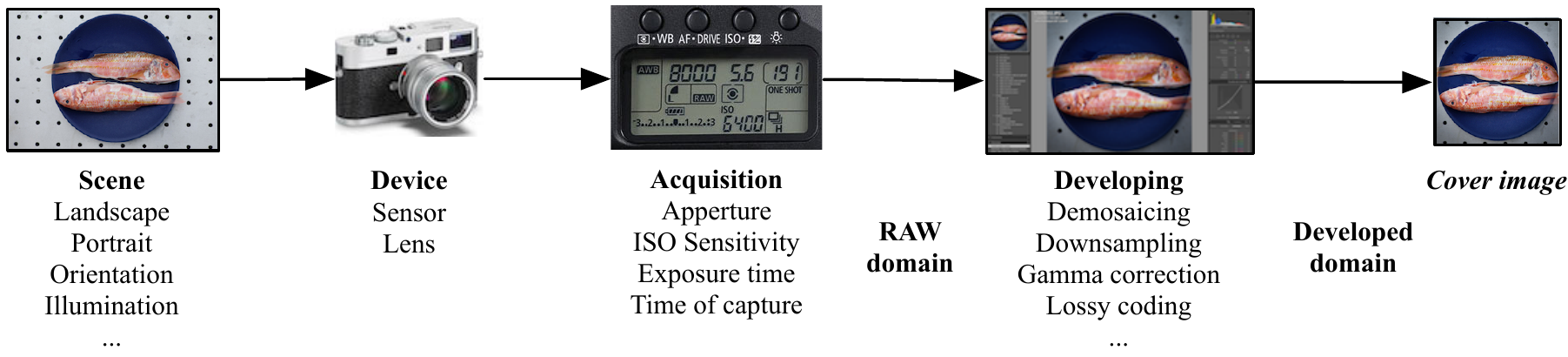}
\par\end{centering}

\caption{Pipeline of the cover image generation process which can be decomposed
into four main steps (scene, device, acquisition, developing) representing
parameters of whole process.\label{fig:Image-acquisition-process}}
\end{figure*}
If the term ``source'' has been first coined with the problem of
``cover-source mismatch'' after the BOSS contest~\cite{bas:hal-00648057,fridrich2011breaking}
in order to denote poor steganalysis performances whenever a steganalyzer
was trained with an image database coming from a set of ``sources''
and tested on another set. In this case the term ``source'' was
associated with a camera device, and other authors~\cite{ker2014steganographer}
have associated a ``source'' with a ``user'' that would upload
a set of pictures on a sharing platform such as FlickR. 

We argue here that a source can be defined w.r.t. the image generation
process depicted in Figure~\ref{fig:Image-acquisition-process} which
shows that the creation of a cover image is linked to the intervention
of different parameters represented by (1) the scene that is captured,
(2) the device which is used, (3) the acquisition settings used during
the capture and (4) the developing step. 

Each parameter is linked with a set of sub-parameters. The scene fluctuates
according to the subject, but also according to the illumination or
the orientation of the camera. The device is composed mainly of two
elements: the sensor (which can be CMOS, CDD, color or monochrome)
and the lens. The acquisition phase relies on three parameters originating
from the device: the lens aperture, the ISO sensitivity and the exposure
time and one parameters which is the time of capture. The developing
step contains lot of processing steps such as gamma correction, demosaicing,
downsampling, JPEG compression, ... it can be done directly within
the camera or using a developing software such as the open-source
dcraw software~\cite{dcraw} or the commercial Lightroom$\copyright$
software. 

Within this paradigm, the classical usage of the term ``source''
as referred in~\cite{bas:hal-00648057,fridrich2011breaking,ker2014steganographer}
represents a setup where the device or more exactly the sensor, is
constant, but a source can also reflect a situation where only a subset
of parameters is constant, such as for example demosaicing algorithm,
the ISO setting, the aperture... Note that in the first case the variations
between images produced from the same source will be extremely large,
even if as shown in~\cite{ker2014steganographer} it will be possible
to separate one source from another, however in the second scenario
the variation between images can be very subtle if only one parameter,
such as for example the time of capture, changes for a given source.
Practically all possible sources are realistic: a casual photographer
using a smartphone will usually adopt 'auto' mode where the acquisition
parameters and the developing step can fluctuate a lot, and a professional
photograph will tend to choose manually his acquisition parameters
(in order for example to minimize the ISO setting) and its developing
process.

\subsection{Steganography via cover-source switching }

The key idea of this paper is to propose a steganographic scheme where
the message embedding will be equivalent to switching from one source
$\mathcal{S}_{1}$ to another source $\mathcal{S}_{2}$; this practically
can be done by designing an embedding that, when applied on $\mathcal{S}_{1}$,
mimics the statistical properties of $\mathcal{S}_{2}$. More specifically
in this paper we have decided to use the sensor noise to model a given
source because its statistical model is rather simple, and to perform
the embedding in such a way that the statistical properties of the
stego images mimic the sensor noise of source $\mathcal{S}_{2}$.
If we refer to Figure~\ref{fig:Image-acquisition-process}, it means
that the only parameters which is fluctuating for this source is the
time of capture. As we shall see in sections~\ref{sec:Sensor-noise-estimation},~\ref{sec:Embedding-with-basic}
and~\ref{sec:Embedding-with-advanced}, the difference between $\mathcal{S}_{1}$
and $\mathcal{S}_{2}$ will stem in the ISO sensitivity. One can argue
that this parameter is reported in the EXIF file of the image, but
EXIF information can be easily edited or even removed using software
such as exiftool~\cite{exifTool}.

Note that this idea of steganography based on mimicking sensor noise
is far from being new. In 1999 Franz and Pfitzmann~\cite{franz1999steganography}
propose a paradigm for a stego-system ``simulating a \emph{usual
process }of data processing'' where the usual process is defined
by the scan process, in this paper the authors study the properties
of scanning noise coming from different scanners. A practical implementation
of this concept in proposed in 2005 by Franz and Schneidewind~\cite{ih-FranzS05},
where the authors model the sensor noise for each pixel by a Normal
distribution and perform the embedding by first estimating the noiseless
scan and secondly adding a noise mimicking the sensor noise. The algorithm
was benchmarked using features derived from wavelet higher order statistics~\cite{holotyak2005blind}
and showed relatively good performance compared with naive noise addition.
It is important to notice that contrary to the work presented here,
if the idea of mimicking the sensor noise is present in~\cite{holotyak2005blind},
it doesn't rely on neither cover-source switching nor a sharp physical
model of the noise in the RAW domain.

Another requirement in order to achieve practical embedding is to
be able to compute the probability of embedding changes $\pi_{k}$
in the developed domain, this in order to perform the practical embedding
but also in order to compute the embedding rate. This particular aspect
will be addressed in sections~\ref{sec:Embedding-with-basic} and~\ref{sec:Embedding-with-advanced}.
Because the embedding scheme relies on the natural statistical noise
of the sensor, we decided to call this steganographic scheme ``Natural
Steganography'' (NS).

The paper is organized as follows: the next section presents the method
which is used to estimate the camera sensor noise of one given cover-source
from raw images, section~\ref{sec:Embedding-with-basic} explains
how to compute the noise mimicking another cover-source in the RAW
domain. When in the developed domain the developing step is basic,
the related embedded payload is also computed. Section~\ref{sec:Embedding-with-advanced}
describes the embedding when the developing step is more elaborated,
three examples are analyzed the gamma correction operation and image
downsampling. Evolutions of the embedded payload w.r.t these operations
are also provided. Finally section~\ref{sec:Experimental-results}
describes the building of a new image database coming from a monochrome
sensor and presents the detectability results related to Natural Steganography.
These results are compared to two state of the art algorithms which
are S-Uniward and S-Uniward using side-information coming from the
conversion from 16-bit to 8-bit. The detectability after gamma correction
or different downsampling schemes is also provided.

\subsection{Notations}

- Capital letters will denote random variables, bold letter denote
vectors or matrices whenever explicitly mentioned. 

- Indexes $(i,j)$ will usually denotes the location of the photo-site\footnote{a photo-site denotes the sensor elementary unit, that after demosaicing
and developing (without geometrical transforms) generates a pixel.} or the pixel in a given image, and the index $k$ denotes a modification
$+k$ on the cover image. Note that these indexes will be omitted
if not necessary for the formula. 

- Notation $\bar{a}$ denotes the version of $a$ after the developing
step. Each element of respectively the raw cover and the raw stego
are denoted $x_{i,j}$ and $y_{i,j}$ and each element of the developed
cover and the developed stego are then denoted $\bar{x}_{i,j}$ and
$\bar{y}_{i,j}$ . The sensor noise is denoted $n_{i,j}$, and the
stego signals in the raw domain and in the developed domain are respectively
denoted $s_{i,j}$ and $\bar{s}_{i,j}$. The virtual noiseless raw
signal is denoted $\mathrm{E}[S_{i,j}]=\mu$. The probability of adding
$k$ on pixel or photo-site $\bar{s}_{i,j}$ is denoted $\pi_{i,j,k}=\mathrm{Pr}[\bar{S}_{i,j}=k]$.

- The 2D convolution between a matrix $\mathbf{m}$ and a filter $\mathbf{f}$
is denoted $\mathbf{m}\star\mathbf{f}$.

\section{Sensor noise estimation and developing pipeline \label{sec:Sensor-noise-estimation}}

We present in this section the different noises affecting the sensor
during a capture and then explain how to estimate the sensor noise.
The last subsection summarizes the image developing pipeline.

\subsection{Sensor noise model\label{sub:Sensor-noise-model} }

Camera sensor noise models have been extensively studied in numerous
publications~\cite{foi2008practical,foi2007noise,european2010standard}
and have already been used in image forensics for camera device identification~\cite{7351518,6662407}.
These models can only be applied to linear sensors such as CDD or
CMOS sensors, but this encompass the majority of modern digital cameras
at the date the paper is written. A camera sensor is decomposed into
a 2D array of photo-sites and the role of each photo-site is to convert
$k_{p}$ photons hitting its surface during the exposure time into
a digit. The conversion involves the quantum efficiency of the sensor
measuring the ratio between $k_{p}$ and the number of charge units
$k_{e}$ accumulated by the photo-site during the exposure time. $k_{e}$
is then converted into a voltage, which is amplified by a gain $K$
(where $K$ is referred as the system overall gain~\cite{european2010standard})
and then quantized.

For each photo-site at location $(i,j)$, the converted signal $x(i,j)$
originates from two components: 
\begin{itemize}
\item The ``dark'' signal $x_{d}(i,j)$ with expectation $\mathrm{E}[X_{d}(i,j)]=\mu_{d}$
which accounts for the number of electrons present without light and
depends on the exposure time and ambient temperature, 
\item The ``electronic'' signal $x_{e}(i,j)$ with expectation $\mathrm{E}[X_{e}(i,j)]=K\mu_{e}$,
which accounts for the number of electrons originating from photons
coming from the scene which is captured. 
\end{itemize}
The expectation $\mu$ of each photo-site response is equal to: 

\begin{equation}
\mu_{i,j}=\mathrm{E}[X(i,j)]=\mathrm{E}[X_{d}(i,j)]+\mathrm{E}[X_{e}(i,j)]=\mu_{d}+K\mu_{e}.\label{eq:mu}
\end{equation}

\begin{figure}
\begin{centering}
\includegraphics{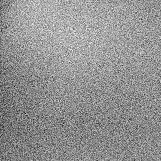}
\par\end{centering}

\caption{Illustration of the sensor noise $n$ on a picture at 2000 ISO captured
with the Leica Monochrome camera. Inactivate the interpolation process
of your pdf viewer for correct rendering. The pixel amplitudes are
scaled for vizualisation purposes. \label{fig:Illustration-of-the}}

\end{figure}

Beside the signal components, there are three types of noise affecting
the acquisition: 
\begin{enumerate}
\item The ``shot noise'' associated with the electronic signal with accounts
for the fluctuation of the number of charge units. Because the electronic
signal comes from the variation of counting events, it has a Poisson
distribution $X_{e}(i,j)\sim\mathcal{P}(\mu_{e})$ and can be approximated
in a continuous setting by a normal distribution $\mathcal{N}(\mu_{e},\sigma_{e}^{2})$
with $\sigma_{e}^{2}=\mu_{e}$, hence the noise associated to the
electronic signal is distributed as $\mathcal{N}(0,\mu_{e})$. Additionally
this noise is independently distributed for each photo-site. An illustration
of the sensor noise is provided in Figure \ref{fig:Illustration-of-the}.
\item The noise related to the ``read-out'' and the amplifier circuit.
The read-out noise associated to the dark signal is independant and
normally distributed as $\mathcal{N}(0,\sigma_{d}^{2})$ and it is
constant for a given camera. 
\item The quantization noise, which is independant and uniformly distributed
with variance $\sigma_{q}^{2}=\Delta^{2}/12$ where $\Delta$ denotes
the quantization step. 
\end{enumerate}
Since these noises are mutually independent, the variance of the sensor
noise can then be expressed as~\cite{european2010standard}:

\begin{equation}
\sigma_{s}^{2}=K^{2}\sigma_{d}^{2}+\sigma_{q}^{2}+K(\mu-\mu_{d}).\label{eq:sigma}
\end{equation}

In the sequel, we make the following approximations for a given cover-source:
we assume that the system gain $K$ is constant for a given $ISO$
setting, that the dark signal is constant with negligible variance
($\sigma_{d}^{2}=0,\mu_{d}=\mathrm{cte_{2}})$, and that the quantization
noise is negligible w.r.t. the shot noise ($\sigma_{q}^{2}=0$). As
we shall see in \ref{sub:Problem-with-8} the two first assumptions
have negligible impact on the performance of the scheme and that the
last assumption does not impact the performance of the steganographic
system whenever 16-bit quantization is considered as side-information.
Finally we also assume that the spacial non-uniformity of the sensor,
which is associated with the photo response non-uniformity (PRNU)
and the dark signal non-uniformity (DSNU), is negligible. 

For a given ISO setting $ISO_{1}$, the global sensor noise $N_{i,j}^{(1)}$
can be approximated using Eq.~(\ref{eq:sigma}) and the above-mentioned
assumptions as normally and independently distributed. We have consequently
a linear relation between the sensor noise variance and the photo-site
expectation $\mu$:

\begin{equation}
N_{i,j}^{(1)}\sim\mathcal{N}(0,a_{1}\mu_{i,j}+b_{1}).\label{eq:dist_sens_noise}
\end{equation}

The acquired photo-site sample $x_{i,j}^{(1)}$ is given by:

\begin{equation}
x_{i,j}^{(1)}=\mu_{i,j}+n_{i,j}^{(1)},\label{eq:photo-site}
\end{equation}

and $X\sim\mathcal{N}(\mu,a_{1}\mu_{i,j}+b_{1})$.

\subsection{Sensor noise estimation\label{sub:Sensor-noise-estimation}}

In order to estimate the model of the sensor noise (i.e. the couple
of parameters $(a,b)$) for a given camera model and a given ISO setting,
we adopt a similar protocol as the one proposed in~\cite{foi2007noise}. 

We first capture a set of $N_{a}$ raw images of a printed photo picturing
a rectangular gradian going from full black to white. The camera is
mounted on a tripod and the light is controlled using a led lightning
system in a dark room. The raw images are then converted to PPM format
(for color sensor) or to PGM format (for B\&W sensor) using the dcraw
open-source software \cite{dcraw} using the command: 

\begin{center}
\texttt{dcraw -k 0 -4 file\_name}
\par\end{center}

which signify that the dark signal is not automatically removed (option\texttt{
-k =0}), and that the captured photo-sites are not post-processed
and plainly converted to 16-bit (option\texttt{ -4}). 

In order to have a process independant of the quantization, the photo-site
outputs are first normalized by dividing them by $y_{max}=2^{16}-1$.
The range of possible outputs is divided into $1/\delta$ segments
of width $\text{\ensuremath{\delta}}$. Each normalized photo-site
location is assigned to one subset of photo-sites $\mathcal{S}_{\ell}$
according to its empirical expectation over the acquired images $\hat{\eta}_{i,j}=\left(\sum_{l=1}^{N_{a}}y_{i,j}^{(l)}/y_{max}\right)/N_{a}$.
The subset index is $\text{\ensuremath{\ell}}=\left[\hat{\eta}_{i,j}/\delta\right]$
where $\left[.\right]$ denotes the integer rounding operation. Once
the segmentation into subsets is performed, the empirical mean is: 

\begin{equation}
\hat{\mu}_{\ell}=\frac{1}{|S_{\ell}|}\sum_{i=1}^{|S_{\ell}|}\mathcal{S}_{\ell}(i),
\end{equation}

where $\mathcal{S}_{\ell}(i)$ denotes the value of a photo-site belonging
to the subset $\mathcal{S}_{\ell}$ and $|.|$ denotes the cardinal
of a set.

The unbiased variance associated to each subset as:

\begin{equation}
\hat{\sigma}_{\ell}^{2}=\frac{1}{|S_{\ell}|-1}\sum_{i=1}^{|S(\ell)|}\left(\mathcal{S}_{\ell}(i)-\hat{\mu}_{\ell}\right)^{2}.
\end{equation}

\begin{figure}[h]
\begin{centering}
\includegraphics[bb=30bp 0bp 526bp 412bp,clip,width=1\columnwidth]{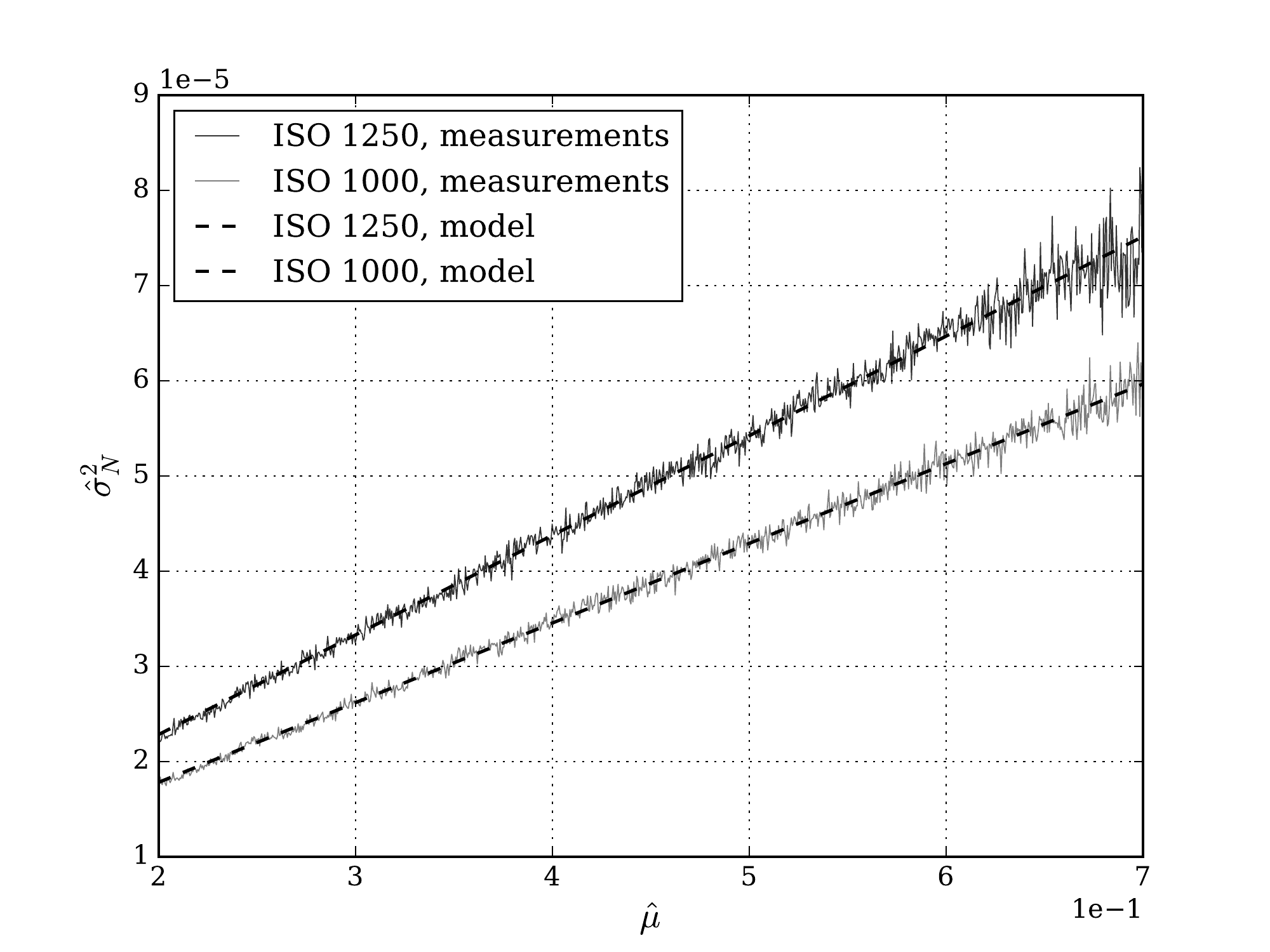}
\par\end{centering}

\caption{Sensor noise estimation for the Leica Monochrome camera and 1000 IS0
and 1250 ISO on normalized images. The estimated coefficients of the
linear model are respectively $(a_{1},b_{1})=(8.36\,10^{-5},1.11\,10^{-6})$
and $(a_{2},b_{2})=(10.46\,10^{-5},1.95\,10^{-6})$ for this setup.\label{fig:Sensor-noise-estimation} }
\end{figure}

As an illustration, Figure~(\ref{fig:Sensor-noise-estimation}) plots
the relation in solid lines between $\hat{\mu}_{\ell}$ and $\hat{\sigma}_{\ell}^{2}$
for $N_{a}=20$ raw images captured with a Leica M Monochrome Type
230 at 1000 ISO and 1250 ISO for $\delta=5\,10^{-5}$. 

The last step consists in estimating the parameters $(\hat{a},\hat{b})$,
this is done by linear regression $\tilde{\sigma}_{N}^{2}=f(\hat{\mu})=\hat{a}\hat{\mu}+\hat{b}$.
We see on the same figure that the linear relation, depicted by the
dashed lines, is rather accurate. 

\subsection{The developing pipeline\label{sub:Image-development-pipeline}}

\begin{figure}[h]
\begin{centering}
\includegraphics[width=1\columnwidth]{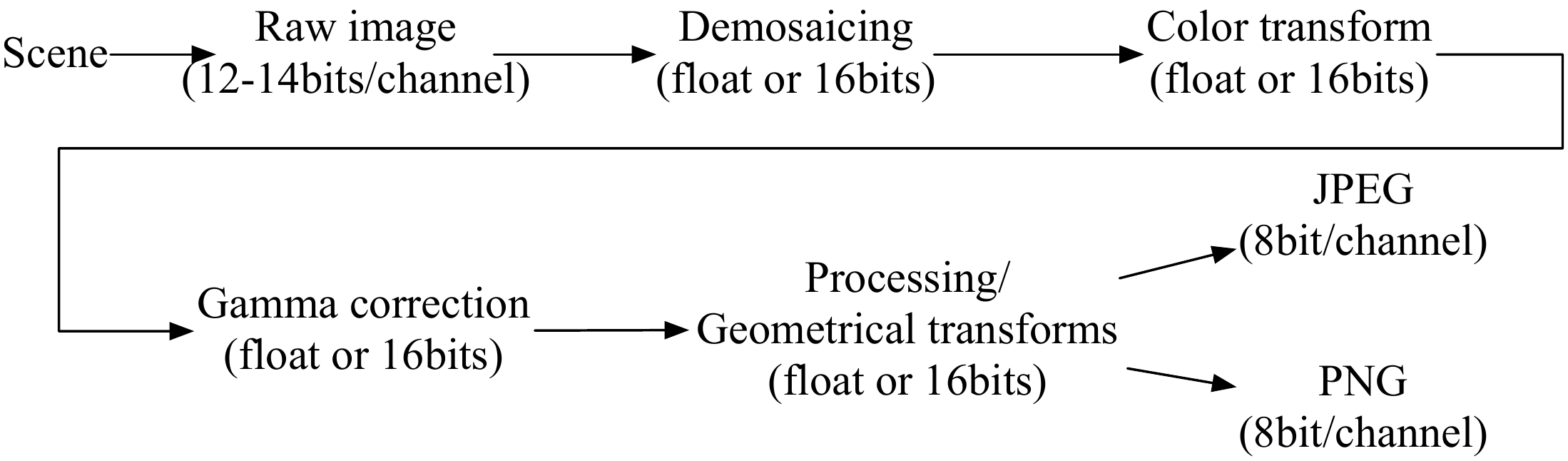}
\par\end{centering}

\caption{The image developing pipeline.\label{fig:The-image-development}}
\end{figure}

Since the goal of any steganographic scheme is to embed message on
a published image and not a raw one, we recall briefly in this subsection
the different steps leading to the generation of a developed picture.
The image developing pipeline, depicted on Figure~\ref{fig:The-image-development}
will be further used in sections~\ref{sec:Embedding-with-basic}
and~\ref{sec:Embedding-with-advanced}. 

In a raw image each photo-site is usually represented using 12 or
14 bits per channel depending on the sensor bit depth, and most of
the color digital cameras only record one color component per photo-site.
Depending of the computational power the following process can be
performed either using either double, float or unsigned integer 16-bit
precision. 

The first step consist in interpolating for each photo-site the two
missing color components via the demosaicing process (see also section
sub:Demosaicing-and-color). This process can be linear or non-linear. 

A color transform is then applied, and converts the original camera
color domain to the reference domain XYZ (this transform depends on
the camera make) and then to a color space such as sRGB, adobe RGB,
wide RGB, ... The white balance correction can also performed at this
stage. All the color conversion operations can be expressed as multiplications
by $3\times3$ matrices and are consequently linear. 

The color components are then corrected using the classical gamma
correction $y=x^{1/\gamma}$ which is a non-linear sample-wise transform. 

The next step encompasses a large set of possible processing operations
and consists in processing the image by applying for example local
contrast enhancement, denoising, adding grain, changing the contrast,
... and applying geometrical transforms such as cropping, image rescaling
(up-sampling or down-sampling), rotation, lens-distortion compensation,
... 

The last step of this pipeline consists in exporting the image in
a lossy compressed format such as JPEG or a lossless format such as
PNG, in each case a quantization to 8-bit/channel is performed. 

Note that depending on developing software which is used, this pipeline
can be completely known whenever the software source code is public,
or unknown if the software is private, but this does not mean for
the last case that the processing operation cannot be reverse-engineered.

\section{Embedding for OOC monochrome pictures \label{sec:Embedding-with-basic}}

\begin{figure}
\begin{centering}
\includegraphics[width=1\columnwidth]{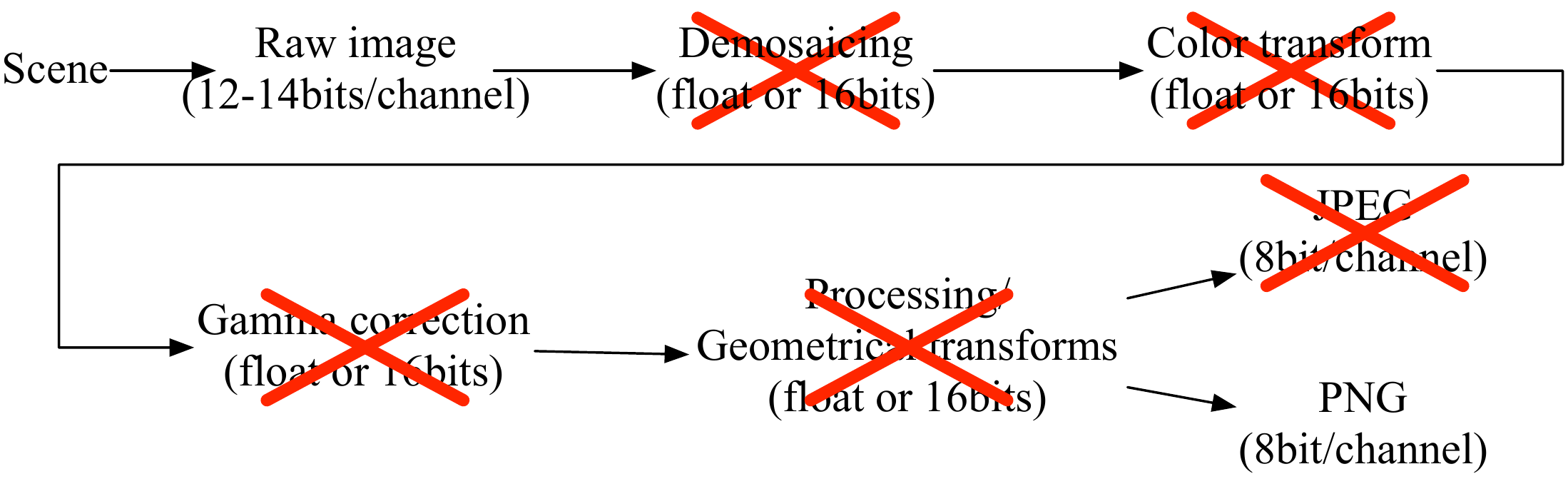}
\par\end{centering}

\caption{The basic developing pipeline associated to a monochrome sensor.\label{fig:The-basic-development}}
\end{figure}

We first propose in this section a steganographic system practically
working for a basic developing setup, this system is realistic for
a monochrome sensor where nor demosaicing neither color transform
is possible. As depicted in Figure~\ref{fig:The-basic-development},
we also assume that the developed images do not undergo gamma correction
or further processing and only suffer 8-bit quantization. We can call
this type of images ``Out Of Camera'' (OOC) Pictures. In the next
section we show how to deal with more advanced developing processes. 

\subsection{Principle of the embedding\label{sub:Principle-of-the}}

We propose to model the stego signal $S_{i,j}$ is such a way that
it mimics the model of images captured at $ISO_{2}>ISO_{1}$. 

Based on the assumptions made in~\ref{sub:Sensor-noise-model}, the
equivalent of (\ref{eq:dist_sens_noise}) and (\ref{eq:photo-site})
for a camera sensitivity parameter $ISO_{2}$ are $N_{i,j}^{(2)}\sim\mathcal{N}(0,a_{2}\mu_{i,j}+b_{2})$
and $x_{i,j}^{(2)}=\mu_{i,j}+n_{i,j}^{(2)}$.

Since the sum of two independent noises normally distributed is normal
with the variances summing up, we can write that $x_{i,j}^{(2)}=\mu_{i,j}+n_{i,j}^{(1)}+s'_{i,j}=x_{i,j}^{(1)}+s'_{i,j}$
with $S'_{i,j}\sim\mathcal{N}(0,(a_{2}-a_{1})\mu_{i,j}+b_{2}-b_{1})$
representing the noise necessary to mimic image captured at $ISO_{2}$. 

Assuming that the observed photo-site is very close to its practical
expectation, i.e. that $\mu_{i,j}\simeq x_{i,j}^{(1)}$, $x{}_{i,j}^{(2)}$
can be approximated by:

\begin{equation}
x_{i,j}^{(2)}\simeq x_{i,j}^{(1)}+s{}_{i,j}\triangleq y_{i,j},\label{eq:cover-source-switch}
\end{equation}

with:

\begin{equation}
S_{i,j}\sim\mathcal{N}(0,(a_{2}-a_{1})x_{i,j}^{(1)}+b_{2}-b_{1}),\label{eq:emb_noise}
\end{equation}

adopting the following notations $a'\triangleq a_{2}-a_{1}$, $b'\triangleq b_{2}-b_{1}$
, $\sigma_{S}^{2}\triangleq a'x_{i,j}^{(1)}+b'$, and the photo-site
of the stego image is distributed as:

\begin{equation}
Y_{i,j}\sim\mathcal{N}(x_{i,j}^{(1)},\sigma_{S}^{2}).\label{eq:Stego-photo-site}
\end{equation}

Note that equation~(\ref{eq:cover-source-switch}) shows explicitly
the principle of cover-source switching which is simply represented
in this case by adding an independant noise on each image photo-site
to generate the stego photo-site $y_{i,j}$. The distribution of the
stego signal in the continuous domain (see~(\ref{eq:emb_noise}))
takes into account the statistical model of the sensor noises estimated
for two ISO settings using the procedure presented in section~\ref{sub:Sensor-noise-model}.

\subsection{16-bit to 8-bit quantization\label{sub:16-bit-to}}

For OOC images, the only developing process lies in the 8-bit quantization
function, consequently the goal here is to compute the embedding changes
probabilities $\pi_{i,j}(k)=\mathrm{Pr}[\bar{S}_{i,j}=k]$ after this
process. 

These probabilities can be either used to simulate optimal embedding,
or cost additive costs $\rho_{i,j}$ can be derived and used to feed
a multilayered Syndrome Trellis Code using the ``flipping lemma''~\cite{filler2011minimizing}
as $\rho_{i,j}=\ln\left(\tilde{\pi}_{i,j}/(1-\tilde{\pi}_{i,j})\right)$
with $\tilde{\pi}_{i,j}=\max\left\{ \pi_{i,j},1-\pi_{i,j}\right\} $
(see also section VI of \cite{filler2011minimizing} for Q-ary embedding
and multi-layered constructions). 

\begin{figure}[h]
\begin{centering}
\includegraphics[bb=0bp 0bp 576bp 360bp,width=1\columnwidth]{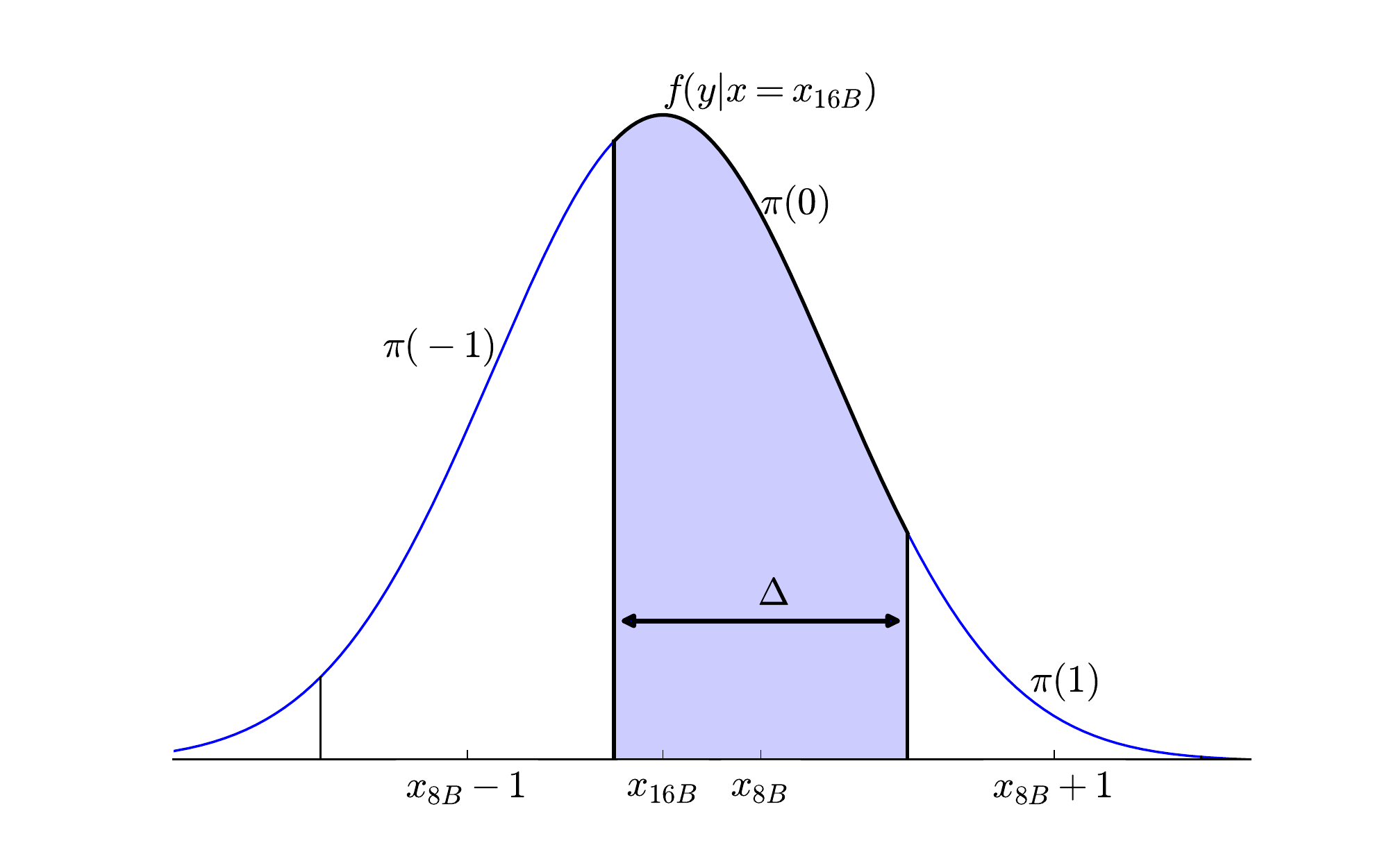}
\par\end{centering}

\caption{Computation of the embedding probabilities after 8-bit quantization\label{fig:Computation-of-the}.}
\end{figure}

We use the high resolution continuous assumption given by~(\ref{eq:pi_k})
and then we compute the discretized probability mass function after
a quantization step of size $\Delta$ (typically $\Delta=256$ by
quantizing from 16-bit resolution to 8-bit resolution). 

The distribution of the stego photo-site $y$ for a given cover photo-site
value coded using 16 bits $x_{16B}$ is depicted in Figure~(\ref{fig:Computation-of-the}). 

The embedding probabilities are directly linked to the 8 bits quantized
value $x_{8B}=Q_{\Delta}(x_{16B})$ - where $Q_{\Delta}(.)$ denotes
the quantization function - and the pdf of the Normal distribution: 

\begin{equation}
\begin{array}{ccc}
\pi(k) & = & \intop_{u_{k}}^{u_{k+1}}f(y|x=x_{16B})dy,\\
 & = & \frac{1}{2}\left(\mathrm{erf}\left(\frac{u_{k+1}-x_{16B}}{2\sigma_{S}^{2}}\right)-\mathrm{erf}\left(\frac{u_{k}-x_{16B}}{2\sigma_{S}^{2}}\right)\right),
\end{array}\label{eq:pi_k}
\end{equation}

with $u_{k}=x_{8B}-(0.5-k)\Delta$..

Once the embedding probabilities are computed for each pixel, it's
possible to derive the payload size using the entropy formula: 

\begin{equation}
H(\overline{S})=-\sum_{i,j,k}\pi_{i,j}(k)\log_{2}\pi_{i,j}(k).\label{eq:Entropy}
\end{equation}

\section{Embedding with advanced developing\label{sec:Embedding-with-advanced}}

\begin{figure*}[h]
\begin{centering}
\begin{tabular}{cccccc}
\includegraphics[width=0.12\paperwidth]{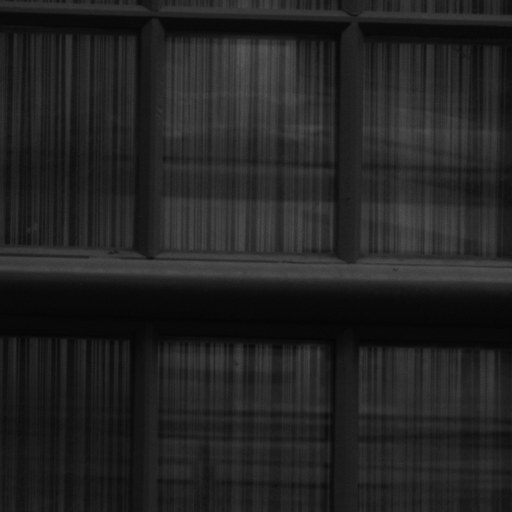} & \includegraphics[width=0.12\paperwidth]{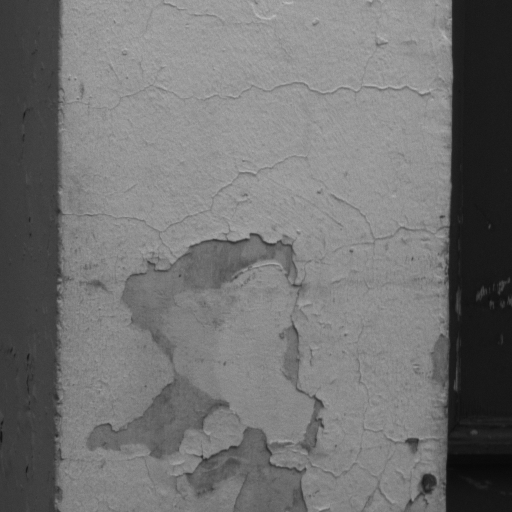} & \includegraphics[width=0.12\paperwidth]{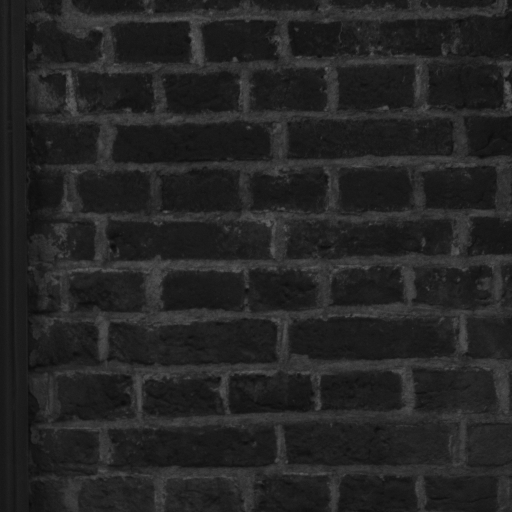} & \includegraphics[width=0.12\paperwidth]{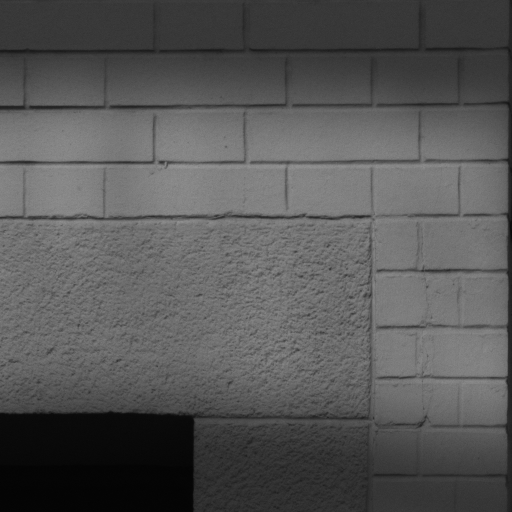} & \includegraphics[width=0.12\paperwidth]{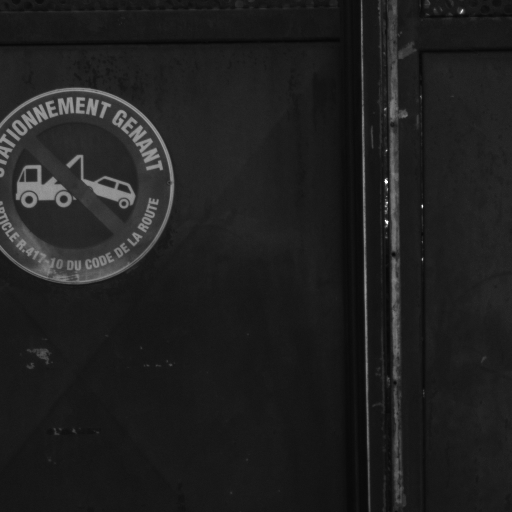} & \includegraphics[width=0.12\paperwidth]{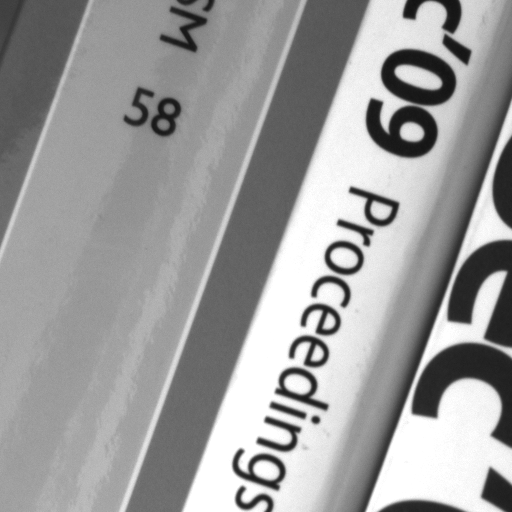}\tabularnewline
\end{tabular}
\par\end{centering}

\caption{Six samples from MonoBase.\label{fig:Five-samples-from}}
\end{figure*}

The main challenge of Natural Steganography is to be able to sample
the stego signal in the developed domain (i.e. $\bar{S}_{i,j}$).
This task can be challenging whenever the developing process is either
unknown or difficult to model. On the contrary whenever the transform
is non-linear but sample-wise, or linear vector-wise, the distribution
of the stego signal $S$ can be computed, possibly using conditional
distributions. In order to give ``proof of concept implementations'',
we focus here on popular processes which are (1) gamma correction,
(2) demosaicing, (3) color transform and (4) down-sampling.

Note that except for the JPEG-compression which is left for further
works, these processes address the main limitations of the embedding
method proposed in section~\ref{sec:Embedding-with-basic}.

\subsection{Gamma correction\label{sub:Gamma-correction}}

The gamma correction is a sample-wise operation defined by $y_{\gamma}\triangleq\Gamma(y)=y_{max}(y/y_{max})^{1/\gamma}$
with $y_{max}=2^{16}-1$, its inverse transform given by $\Gamma^{-1}(y)=y_{max}(y_{\gamma}/y_{max})^{\gamma}$. 

In order to compute the distribution of the stego signal after gamma
correction, one can simply compute the distribution of the transform
of a continuous variable as:

\begin{equation}
\begin{array}{ccc}
f_{Y_{\gamma}}(y_{\gamma}) & = & f_{Y}(y_{\gamma})\frac{d}{dy}\Gamma^{-1}(y),\\
 & = & \frac{1}{\sqrt{2\sigma_{S}^{2}\pi}}\mathrm{exp}\left(-\frac{\left(y_{\gamma}-x^{(1)}\right)^{2}}{2\sigma_{S}^{2}}\right)\gamma(y_{\gamma}/y_{max})^{\gamma-1}.
\end{array}
\end{equation}

However, since in practice $\sigma_{S}\ll x^{(1)}$ we can use a first
order taylor expansion of the gamma correction, given by: 

\begin{equation}
\begin{array}{ccc}
y_{\gamma} & \simeq & \Gamma(x^{(1)})+(y-x^{(1)})(x^{(1)}/y_{max})^{1/\gamma-1}/\gamma,\\
 & = & \Gamma(x^{(1)})+\alpha s,
\end{array}
\end{equation}

with $\alpha=(x^{(1)}/y_{max})^{1/\gamma-1}/\gamma$. This means that
the gamma correction acts as an affine transform on the stego signal. 

Consequently, as a first approximation, the stego signal $S_{\gamma}$
after gamma correction can be considered as normally distributed:

\begin{equation}
S_{\gamma}\sim\mathcal{N}(0,\alpha^{2}\sigma_{S}^{2}),\label{eq:emb_noise-gamma}
\end{equation}

and the distribution of the stego photo-site is given by: 

\begin{equation}
Y_{\gamma}\sim\mathcal{N}(z_{\gamma}(x^{(1)}),\alpha^{2}\sigma_{S}^{2}).\label{eq:Stego-photo-site-gamma}
\end{equation}
Because gamma correction is a sample-wise operation, the stego signal
is independently distributed, and the embedding probabilities after
8-bit quantization can be directly computed as:

\begin{equation}
\begin{array}{ccc}
\pi_{\gamma}(k) & = & \intop_{u'_{k}}^{u'_{k+1}}f(y_{\gamma}|x=\Gamma(x_{16B}))dy,\\
 & = & \frac{1}{2}\left(\mathrm{erf}\left(\frac{u'_{k+1}-\Gamma(x_{16B})}{2\alpha^{2}\sigma_{S}^{2}}\right)-\mathrm{erf}\left(\frac{u'_{k}-\Gamma(x_{16B})}{2\alpha^{2}\sigma_{S}^{2}}\right)\right),
\end{array}\label{eq:pi_k_gamma}
\end{equation}

with $u'_{k}=x'_{8B}-(0.5-k)\Delta$, $x'_{8B}=Q_{\Delta}(\Gamma(x_{16B}))$.
The payload size is consequently given as $H(\overline{S}_{\gamma})=-\sum_{i,j,k}\pi_{\gamma,i,j}(k)\log_{2}\pi_{\gamma,i,j}(k)$.

\subsection{Demosaicing and color images\label{sub:Demosaicing-and-color}}

The goal of demosaicing is to predict for each photo-site the two
missing color channels from neighboring photo-sites that, through
the Color Filter Array, record other channels. Popular demosaicing
schemes include bi-linear filtering which use linear interpolation
from the neighbors of the same channel, or Variable Number of Gradients
(VNG)~\cite{chang1999color}, Patterned Pixel Grouping (PPG)~\cite{PPG},
Adaptive Homogeneity-Directed (AHD)~\cite{hirakawa2005adaptive},
which are more advanced schemes using the correlation between the
channels to predict the one recorded by the photo-site but also edge-directionality
and color homogeneity. Note that the implementation of these algorithms
is available in dcraw or the library libraw~\cite{libraw}. One notable
point of the demosaicing procedure is that the recorded photo-sites
values stay unchanged and the sensor noise is still distributed as~(\ref{eq:dist_sens_noise})
for unpredicted values. 

Without loss of generality we assume that there is no other operations
in the developing process than demosaicing, we can consequently have
a straightforward implementation of NS for color images by (1) embedding
the message on the photo-site values using the algorithm proposed
in section~\ref{sec:Embedding-with-basic} and (2) performing the
demosaicing process to generate the Stego image. The message will
be then be decoded only from the non-interpolated values. 

One important consequence is that the embedding ratio for a given
image using NS will only depends on the recorded sensor value by the
camera and will be the same whether the sensor uses a Color Filter
Array and produce a color image or is a black and white sensor and
produces gray-level pictures. It is not because the image is recorded
using three channel that the embedded payload is three time bigger.
Note that this payload increase for color images is however possible
for color camera sensors such as Foveon sensors which acquire the
three channels for each photo-site~\cite{foveon}.

\subsection{Color transform\label{sub:Color-transform}}

The color transform is a linear operation and consequently the 3 components
of the stego signal after demosaicing $[s_{CR},s_{CG},s_{CB}]^{T}$
can be expressed as:

\begin{equation}
\left[\begin{array}{c}
s_{R}\\
s_{G}\\
s_{B}
\end{array}\right]=\left[\begin{array}{ccc}
c_{11} & c_{12} & c_{13}\\
c_{21} & c_{22} & c_{23}\\
c_{31} & c_{32} & c_{33}
\end{array}\right]\left[\begin{array}{c}
s_{CR}\\
s_{CG}\\
s_{CB}
\end{array}\right],\label{eq:color_transform}
\end{equation}

where $[s_{R},s_{G},s_{B}]^{T}$ represents the RGB vector after the
color transform. In the case the color transform does only white balance
($a_{i,j}=0$ for $i\neq j$) we can adopt the same strategy as in
section~(\ref{sub:Demosaicing-and-color}) with $S_{R}\sim\mathcal{N}(0,\sigma_{R}^{2}\triangleq c_{11}^{2}\sigma_{S_{CR}}^{2})$,
$S_{G}\sim\mathcal{N}(0,\sigma_{G}^{2}\triangleq c_{22}^{2}\sigma_{S_{SG}}^{2})$,
$S_{B}\sim\mathcal{N}(0,c_{B}^{2}\triangleq c_{33}^{2}\sigma_{S_{SB}}^{2})$.
The embedding probabilities and message length computed using~(\ref{eq:pi_k})
and~(\ref{eq:Entropy}) using $\sigma_{S}^{2}=\sigma_{S\{R,G,B\}}^{2}$.

For a classical color transform, we have to proceed differently and
we propose here a sub-optimal scheme that will embed a payload only
on half on the pixels\footnote{This is because the information is carried on green photo-site which
on a bayer CFA represents half of the photo-sites.} and for demosaicing that predicts components only from photo-sites
coding the same component (like bilinear demoisaicing): 
\begin{enumerate}
\item We start by adding a noise distributed as the stego signal on the
photo-sites recording the blue and red channels. This noise is not
used to convey a message and only enable cover-source switching.
\item We interpolate the blue and red components of the sensor noises $s_{CR}$
and $s_{SB}$ for all photo-sites recording the green channel using
demosaicing.
\item We have then $s_{R}=\mathrm{cte}_{1}+c_{12}s_{CG}$, $s_{G}=\mathrm{cte}_{2}+c_{22}s_{CG}$
, $s_{B}=\mathrm{cte}_{3}+c_{32}s_{CG}$ and we select the component
$C_{max}\in\{R,G,B\}$ associated to the highest $c_{max}=c_{i2}$
to carry the payload. Usually it is also the green component of the
new space. We do so in order to maximize the embedding capacity of
the scheme.
\item We compute the embedding probabilities using~(\ref{eq:pi_k}) with
the appropriate $\sigma_{S'}^{2}=c_{max}^{2}\sigma_{Cmax}^{2}$ and
we modify the component accordingly, embedding the payload using STCs
or simulating embedding.
\item We draw a random variable distributed according to the portion of
the gaussian distribution where k is selected in the previous step
(see figure~(\ref{fig:Computation-of-the})), 
\item We compute the modifications on the two other components ($\{(R,G,B)-C_{max}\}$)
by quantizing the random-variable according to the resolution.
\item We interpolate the other green raw component using demosaicing and
we perform the color transform on photo-sites coding R and B.
\item The payload is decoded by reading the values of the color channel
$C$ on developed photos-sites encoding the green information.
\end{enumerate}
It is important to notice that this embedding process will bring a
positive correlation between color components whenever $c_{max}$
and the $c_{i2}$ of the selected component are of same sign, and
a negative correlation otherwise. Note that the idea of forcing correlation
between embedding changes has already empirically been proposed in
a variation of CMD for color images~\cite{tang2016clustering}, we
bring here a more theoretical insight of why this is necessary.

\subsection{Down-sampling (and up-sampling)\label{sub:Down-sampling-(and-up-sampling)}}

We propose in this subsection strategies to deal with image down-sampling.
We restrict our analysis to integer down-scaling factors $c\in\mathcal{\mathbb{N}}^{+}$. 

Note that upscaling with integer factors is the similar strategy than
the one of demosaiced images since the predicted pixels are constructed
from the non-interpolated ones and do not carry any information. As
a consequence an up-sampled image carrie the same payload than the
original one.

For down-sampling we distinguish three strategies: sub-sampling, box
down-sampling and down-sampling using convolutional kernels such as
tent down-sampling.\\

\subsubsection{Sub-sampling\label{sub:Sub-sampling}}

Sub-sampling consists in selecting pixels distant by $kc$ pixels
($k\in\mathcal{\mathbb{N}}^{+}$) on each column and row of the image.
For a stationary image, naïve sub-sampling consequently does not modify
the average embedding rate, but this sub-sampling method is rarely
used in practice since it creates aliasing.\\

\subsubsection{Box down-sampling\label{sub:Box-down-sampling}}

\begin{figure}[h]
\begin{centering}
\begin{tabular}{ccc}
\includegraphics[width=0.3\columnwidth]{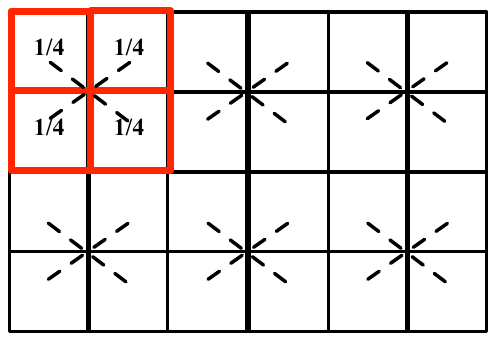} & \includegraphics[width=0.35\columnwidth]{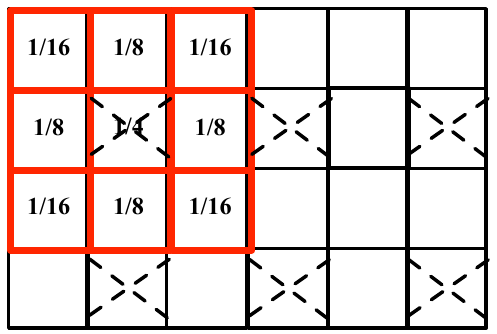} & \includegraphics[width=0.15\columnwidth]{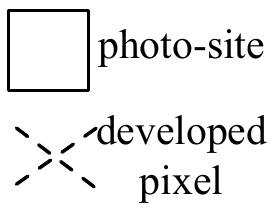}\tabularnewline
(a) & (b) & \tabularnewline
\end{tabular}
\par\end{centering}

\caption{Neighborhood and filters for box down-sampling (a) and tent-down-sampling
(b) for a factor 2, crosses represent the location of the predicted
pixel. \label{fig:Neighborhood-and-filters}}
\end{figure}

Box down-sampling consists in computing the averages of disjoint blocs
of size $c\times c$ to compute down-sampled values (see Figure~(\ref{fig:Neighborhood-and-filters})(a)).
The stego signal is consequently averaged on $c^{2}$ pixels it is
distributed as: 
\begin{equation}
S_{box}\sim\mathcal{N}\left(0,\sigma_{box}^{2}\right),\label{eq:emb_down_box}
\end{equation}

with $\sigma_{box}^{2}=\left[\sum_{i=1}^{c^{2}}\sigma_{S}^{2}(i)\right]/c^{4}=(a'\overline{x}+b')/c^{2}.$

This means that on the developed image, the variance of the stego
signal is divided by $c^{2}$. Since the blocs are disjoint the noise
is still independently distributed and formulas~(\ref{eq:pi_k})
and (\ref{eq:Entropy}) can be used in order to compute the embedding
rate. Equation \ref{eq:emb_down_box} is equivalent to write that 

As it will be analyzed in section~\ref{sub:Downsampling-and-scaling-res},
we can already notice that the embedding rate is a decreasing function
of the scaling factor in this case.\\

\subsubsection{Tent down-sampling\label{sub:Tent-down-sampling}}

We now investigate more elaborated filters and particularly the problem
of tent down-sampling (aka triangle down-sampling or bilinear down-sampling).
This analysis enables to understand how it is possible to embed a
message in the downscaled image using this particular filter, but
it can also be adapted to all the class of linear filters, including
for example the Gaussian kernel or the Lanczos kernel~\cite{turkowski1990filters}. 

\begin{figure}[h]
\begin{centering}
\begin{tabular}{c}
\includegraphics[width=0.5\columnwidth]{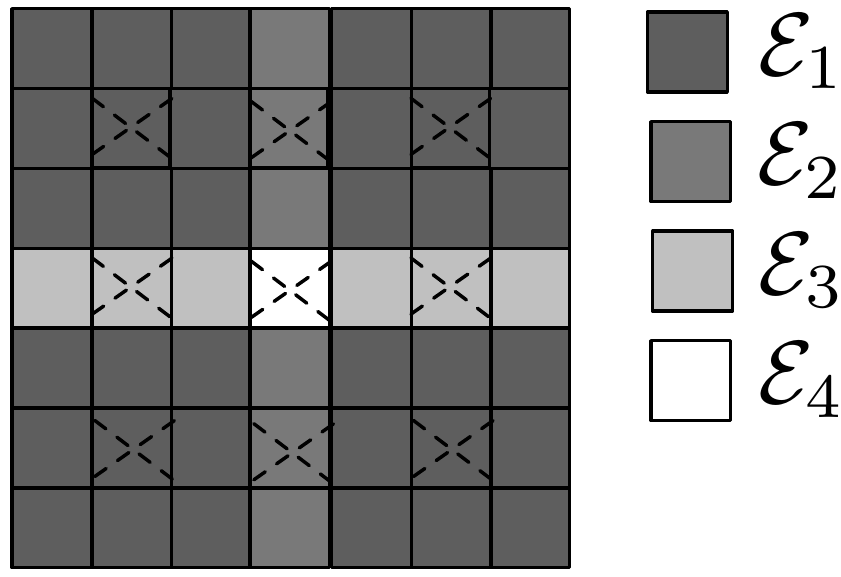}\tabularnewline
\end{tabular}
\par\end{centering}

\caption{Embedding steps used for tent-down sampling: dark grey photo-sites
are sampled during the first step, light grey photo-sites during the
second step and white photo-sites during the last step. \label{fig:Tent-steps}}
\end{figure}

As for color transform, we propose an embedding scheme that we believe
is sub-optimal, i.e. that does't convey the maximum payload but that
ensures that the stego signal mimics correctly the sensor-noise. Without
loss of generality the principle of the embedding scheme is explained
for $c=2$ and in this case the tent filter and down-sampling process
is illustrated on Figure~\ref{fig:Neighborhood-and-filters}(b). 

This method starts by decomposing the down-scaled pixels into four
disjoint lattices and associated subsets. Note that the embedding
strategy is then similar to the ones designed to unable synchronization
between embedding changes such as the ``CMD'' strategy or the ``Sync''
strategy~\cite{denemark2014selection,li2015strategy}.

This decomposition is done in order to obtain developed pixels for
which the stego signal is independently distributed conditionally
to the neighborhood. On figure~\ref{fig:Neighborhood-and-filters}(b)
we can see that the first and second developed pixels (represented
by crosses) are not independant since 3 photo-sites contribute to
the computation of both pixels, on the contrary the first and third
pixels of the first row are independant.

The embedding procedure can be decomposed into four steps, each of
them associated to a given subset of photo-sites.

In this section we adopt the following notations: indexes $i,j$ are
centered on the pixel to develop ($\{-1,0,1\}$ represents respectively
$\{1,2,3\}$ rows or columns), and the tent filter is denoted as a
$3\times3$ matrix with coefficient $c_{i,j}$. The four different
subsets $\{\mathcal{E}_{1},\mathcal{E}_{2},\mathcal{E}_{3},\mathcal{E}_{4}\}$
of photo-sites are represented on Figure~\ref{fig:Tent-steps}. Moreover
$\overset{\uparrow}{s}$, $\overset{\downarrow}{s}$, $\overset{\leftarrow}{s}$
and $\overset{\rightarrow}{s}$ denotes stego signal added on the
photo-sites related to neighboring developed pixels according to the
$\uparrow,\downarrow,\leftarrow,\rightarrow$ directions. As an example
it means that $s_{-1,0}=\overset{\uparrow}{s_{1,0}}$. 

The embedding is sequentially performed in 4 steps: 
\begin{enumerate}
\item The first step embeds part of the message (or generate the stego signal)
into pixels belonging to $\text{\ensuremath{\mathcal{E}}}_{1}$. Because
the subset $\mathcal{E}_{1}$ generates independent pixels, the stego
signal in the developed domain is distributed as:

\begin{equation}
\mathcal{N}(0,\sigma_{S1}^{2}),
\end{equation}

with $\sigma_{S1}^{2}=\sum_{i,j=-1}^{1}c_{i,j}^{2}\sigma_{S}^{2}(i,j)$.
Applying results of~\ref{sec:Embedding-with-basic}, one can compute
the embedding probabilities, and the associated payload length associated
to the pixels belonging to $\mathcal{E}_{1}$. In order to be able
to sample the neighboring pixels, once an embedding change is done,
we draw realizations of the 9 underlying photo-sites. This can be
done by computing conditional probabilities or performing rejection
sampling. 

\item Developed pixels belonging to $\text{\ensuremath{\mathcal{E}}}_{2}$
have a sensor noise distributed according to the conditional density
$f(\bar{s}|\overset{\leftarrow}{s}_{i-1,1},\overset{\leftarrow}{s}_{i,1},\overset{\leftarrow}{s}_{i+1,1},\overset{\rightarrow}{s}_{i-1,1},\overset{\rightarrow}{s}_{i,1},\overset{\rightarrow}{s}_{i+1,1})$
and consequently can be expressed as:

\begin{equation}
\mathcal{N}(\mu_{S2},\sigma_{S2}^{2}),
\end{equation}

with $\mu_{S2}=\sum_{i=-1}^{1}c_{i,1}\overset{\leftarrow}{s}_{i,1}+\sum_{i=-1}^{1}c_{i,-1}\overset{\rightarrow}{s}_{i,-1}$
and $\sigma_{S2}^{2}=\sum_{i=-1}^{1}c_{i,0}^{2}\sigma_{S}^{2}(i,0)$.
As for the first step, we can compute embedding probabilities and
payload length for this subset. We can also draw the realizations
of stego signal related to the 3 photo-sites belonging to this subset.
Note that the same applies for steps 3 and 4. 

\item Similarly pixels belonging to $\text{\ensuremath{\mathcal{E}}}_{3}$
have a sensor noise are distributed as:

\begin{equation}
\mathcal{N}(\mu_{S3},\sigma_{S3}^{2}),
\end{equation}

with $\mu_{S3}=\sum_{j=-1}^{1}c_{1,j}\overset{\uparrow}{s}_{1,j}+\sum_{j=-1}^{1}c_{1,j}\overset{\downarrow}{s}_{1,j}$
and $\sigma_{S3}^{2}=\sum_{j=-1}^{1}c_{0,j}^{2}\sigma_{S}^{2}(0,j)$,
and as for step 2, it is possible to draw realizations of the stego
signal. 

\item An finally, pixels belonging to $\text{\ensuremath{\mathcal{E}}}_{4}$
have a sensor noise distributed as:

\begin{equation}
\mathcal{N}(\mu_{S4},\sigma_{S4}^{2}),
\end{equation}

with $\mu_{S4}=\sum_{j=-1}^{1}c_{1,j}\overset{\uparrow}{s}_{1,j}+\sum_{j=-1}^{1}c_{-1,j}\overset{\downarrow}{s}_{-1,j}+c_{0,1}\overset{\leftarrow}{s}_{0,1}+c_{0,-1}\overset{\rightarrow}{s}_{0,-1}$
and $\sigma_{S4}^{2}=c_{0,0}^{2}\sigma_{S}^{2}(0,0)$. For this last
step, notice that only one photo-site is drawn.

\end{enumerate}
Note that since $H(\bar{S}|\overset{\leftarrow}{S}_{i-1,1},\overset{\leftarrow}{S}_{i,1},\overset{\leftarrow}{S}_{i+1,1},\overset{\rightarrow}{S}_{i-1,1},\overset{\rightarrow}{S}_{i,1},\overset{\rightarrow}{S}_{i+1,1})\leq H(\bar{S})$,
the payload length embedded during steps 4 is smaller than the payload
length embedded during 2 and 3, which is smaller than the payload
length embedded during step 1.

\section{Experimental results\label{sec:Experimental-results}}

The goal of this section is to benchmark the detectability of NS,
to compare it with other steganographic schemes using same embedding
payload, but also to analyze the effects of developing operations
w.r.t. both detectability and embedding rates.

\subsection{Generation of ``MonoBase''\label{sub:Generation-of-Mono-Base}}

In order to benchmark the concept of embedding using cover-source
switching, we needed to acquire different sources providing OOC images.
To do so we conducted the following experiment: using a Leica M Monochrome
Type 230 camera, we captured two sets of 172 pictures taken at 1000
ISO or 1250 ISO. In order to have large diversity of contents most
of the pictures were captured using a 21mm lens in a urban environment,
or a 90mm lens capturing cluttered places. 

The exposure time was set to automatic, with exposure compensation
set to -1 in order to prevent over-exposure. A tripod was used so
that pictures for the two sensitivity settings correspond to the same
scene. Each RAW picture was then converted into a 16-bit PGM picture
using the same conversion operation as the one presented in section~\ref{sub:Sensor-noise-estimation}
and each $5212\times3472$ picture was then cropped into $6\times10=60$
PGM pictures of size $512\times512$ to obtain two sets of 10320 16-bit
PGM pictures. We consequently end up with a database of a similar
size than BOSSBase, with pictures of same size that contrary to un-cropped
pictures can be quickly processed either for embedding or feature
extraction. This database called MonoBase is downloadable here~\cite{Monobase}.
Figure~\ref{fig:Five-samples-from} shows several images of MonoBase.

\subsection{Benchmark setup\label{sub:Benchmark-setup}}

For all the following experiments, we adopt the Spatial Rich Model
feature sets~\cite{fridrich2012rich} combined with the Ensemble
Classifier (EC)~\cite{kodovsky2012ensemble} and we report the average
total error $P_{E}=\min((P_{FA}+P_{MD})/2)$ obtained after training
the EC on 10 different training/testing sets divided in 50/50. The
stego database consists of images captured at 1000 ISO perturbed with
an embedding noise mimicking 1250 ISO, and the cover database consists
of images captured at 1250 ISO. In order to have an effect equivalent
with the principle of training using pairs of cover and stego images,
the pairs are constructed using one couple of images capturing the
same scene.

The parameters of the stego signal are denoted $a"$ and $b"$ with
the relations $a"=a'(2^{N_{b}}-1)$ and $b"=b'(2^{N_{b}}-1)^{2}$,
where $a'$ and $b'$ are computed using normalized image values in
order to be resolution independant (see section~\ref{sub:Principle-of-the}).
$N_{b}=16$ when the cover image coded in 16-bit is used and $N_{b}=8$
when the stego image is directly generated from the 8-bit representation
of the cover image.

\subsection{Basic developing and comparison with S-Uniward\label{sub:Basic-developing-and}}

We first benchmark the scheme proposed in section~\ref{sec:Embedding-with-basic}
and generate 8-bit stego images where the stego signal is generated
according to the embedding probabilities computed in~\ref{eq:pi_k}.
Like all the modern steganographic schemes, we forbid embeddings by
attributing wet pixels to cover pixels saturated at 0 or 255. We also
propose a variation of NS where the dark pixels, i.e. the pixels of
the cover have the lowest value after 8-bits quantization, are also
wet. This strategy, even if developed independently, is similar to
the one recently proposed in~\cite{fridrich_icip_16}. For NS, we
used the same values that the ones estimated in section~\ref{sec:Sensor-noise-estimation},
i.e. $a"=2.1\,10^{-5}$ and $b"=8.4\,10^{-7}$.

The two first columns of table~\ref{tab:Results-and-comparison}
show the high undetectability of the proposed scheme, and the small
improvement associated to wet the dark pixels. We note that we are
still around $7\%$ from random guessing, and we think that it can
be due to the different assumption presented in section~\ref{sec:Embedding-with-basic},
particularly the fact that the quantization noise is ignored.

\begin{table}[h]
\begin{centering}
\begin{tabular}{|c||c|c|c|c|c|}
\hline 
 & NS & NS & SUni-SI & SUni & 1000 ISO \tabularnewline
 & wo wet dark &  & 1000 ISO & 1000 ISO & vs 1250 ISO\tabularnewline
\hline 
$P_{E}$ & 41.0\% & \textbf{42.8\%} & 18.2\% & 12.3\% & 26.0\%\tabularnewline
\hline 
\end{tabular}
\par\end{centering}

\caption{Results and comparison with S-Uniward on MonoBase 1000 ISO.\label{tab:Results-and-comparison}}
\end{table}

\begin{figure}[h]
\centering{}%
\begin{tabular}{c}
\includegraphics[width=0.9\columnwidth]{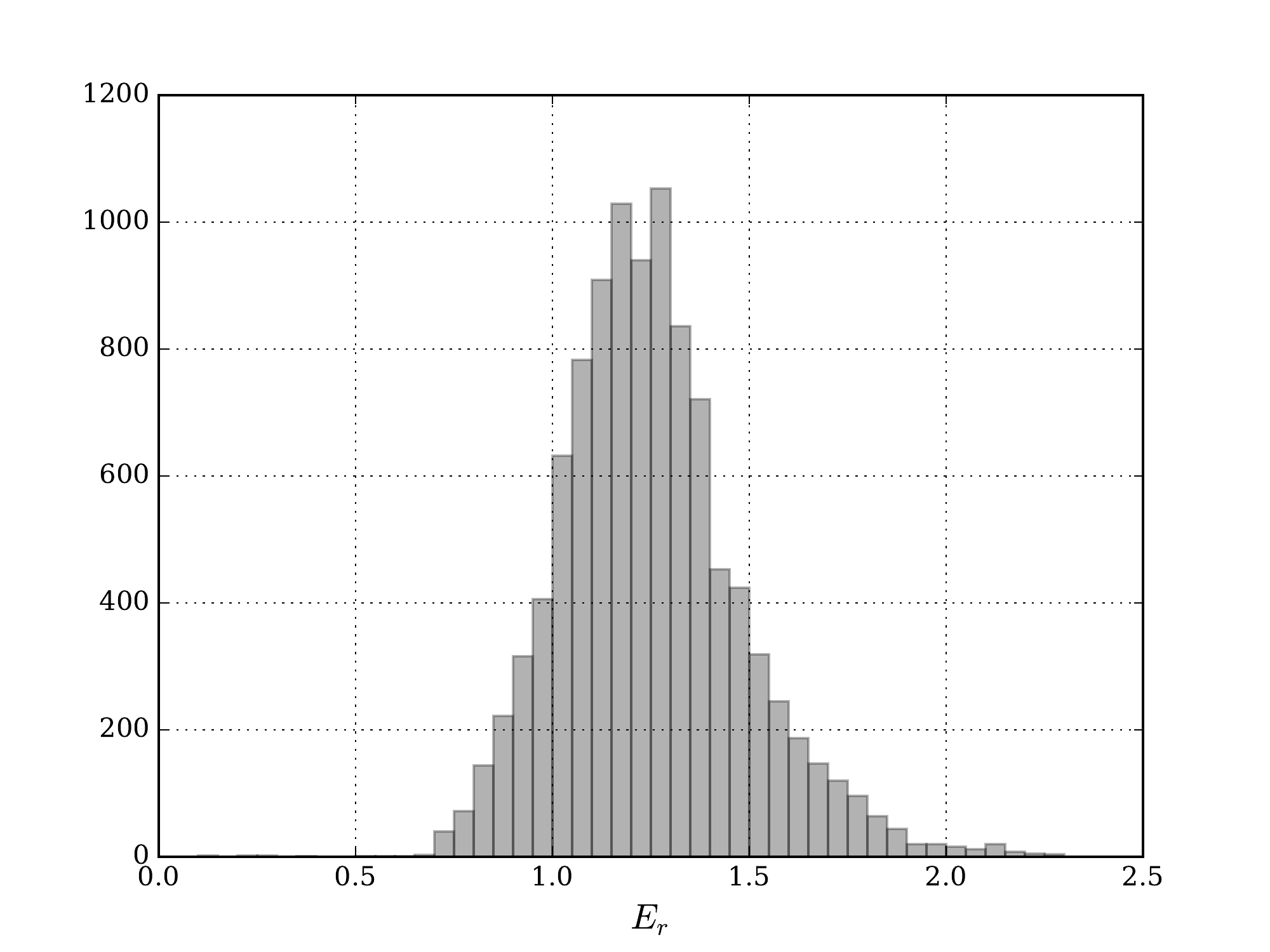}\tabularnewline
\end{tabular}\caption{Histogram of the embedding rates for MonoBase with covers-source switching
from 1000 ISO to 1250 ISO. $\mathrm{E}[E_{r}]=1.24\,\mathrm{pbb}$.
\label{fig:ER_hist}.}
\end{figure}

Figure~\ref{fig:ER_hist} depicts the histogram of the embedding
rate ($E_{r}$) on MonoBase. We can see that most of the embedding
rates are relatively high for steganography with an average of 1.24
bpp for this base. Note that on MonoBase, most of the images are under-exposed,
which means that the average embedding rate should higher for a ``typical''
database. It is important to point here that contrary to most of the
steganographic schemes, the current implementation of NS does not
enable an embedding at a constant payload, but this as the price of
high undetectability. 

The two next columns of Figure~\ref{fig:ER_hist} compare the performances
of NS with S-Uniward. We choose this steganographic scheme because
of its excellent performances and because it has recently been tuned
to take into account the side-information offered by 16-bit to 8-bit
conversion~\cite{denemark-wifs2016}. The two implementations of
S-Uniward where benchmarked on MonoBase 1000 ISO on a fixed embedding
rate equal to the average embedding rate of NS, but we obtained similar
performances on MonoBase 1250 ISO. This scheme, even if excellent
at deriving embedding costs and targeting textured regions of the
image, cannot compete with an embedding scheme which is model-based
and which enable cover-source switching like NS. 

The last column compare our steganalysis task with the classification
task of separating images captured at 1000 ISO from image captured
at 1250 ISO. We can see that this task using the SRM is not an easy
task since the error probability is still rather important.

\begin{figure}
\begin{centering}
\begin{tabular}{c}
\includegraphics[width=0.9\columnwidth]{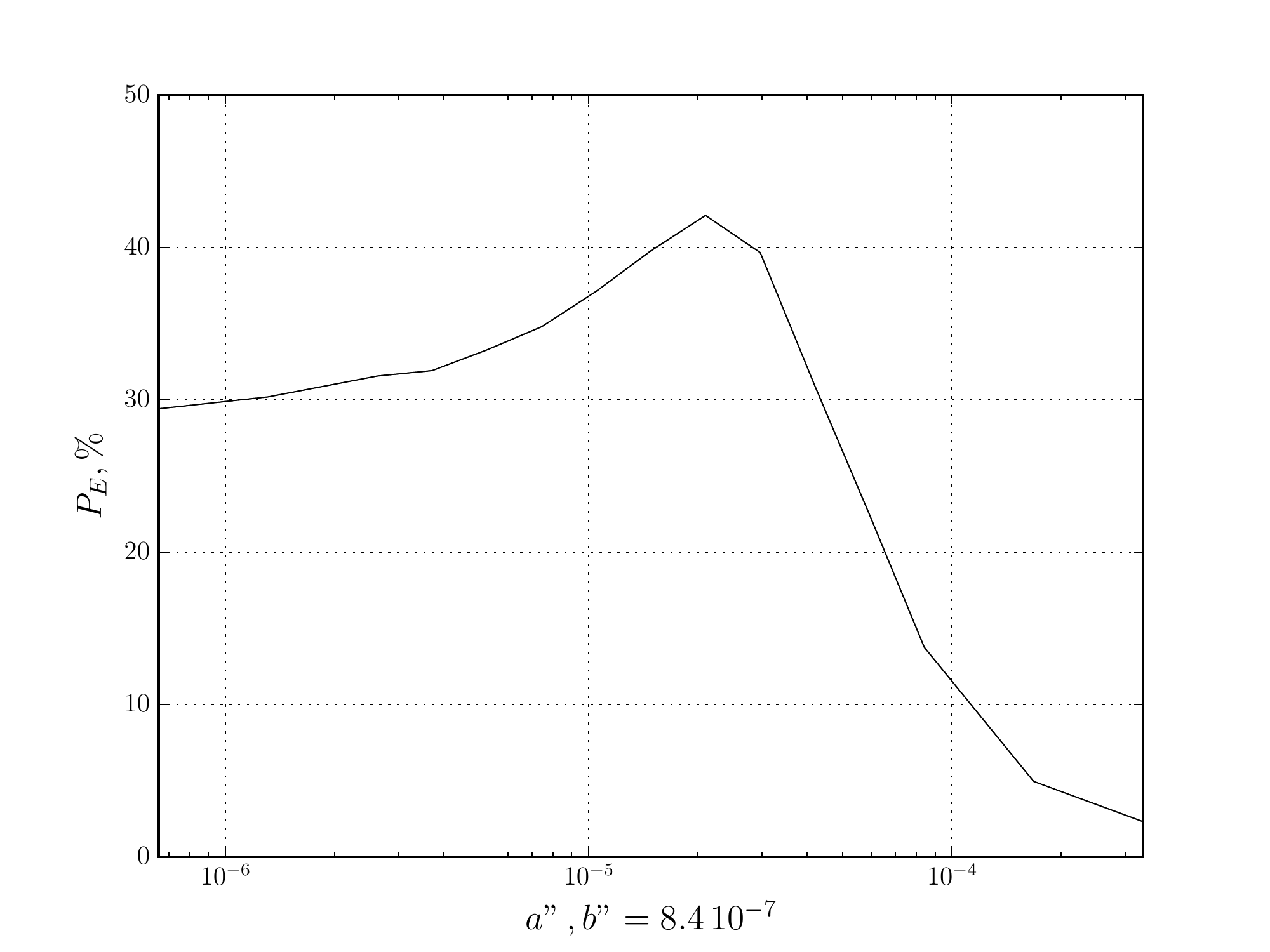}\tabularnewline
\includegraphics[width=0.9\columnwidth]{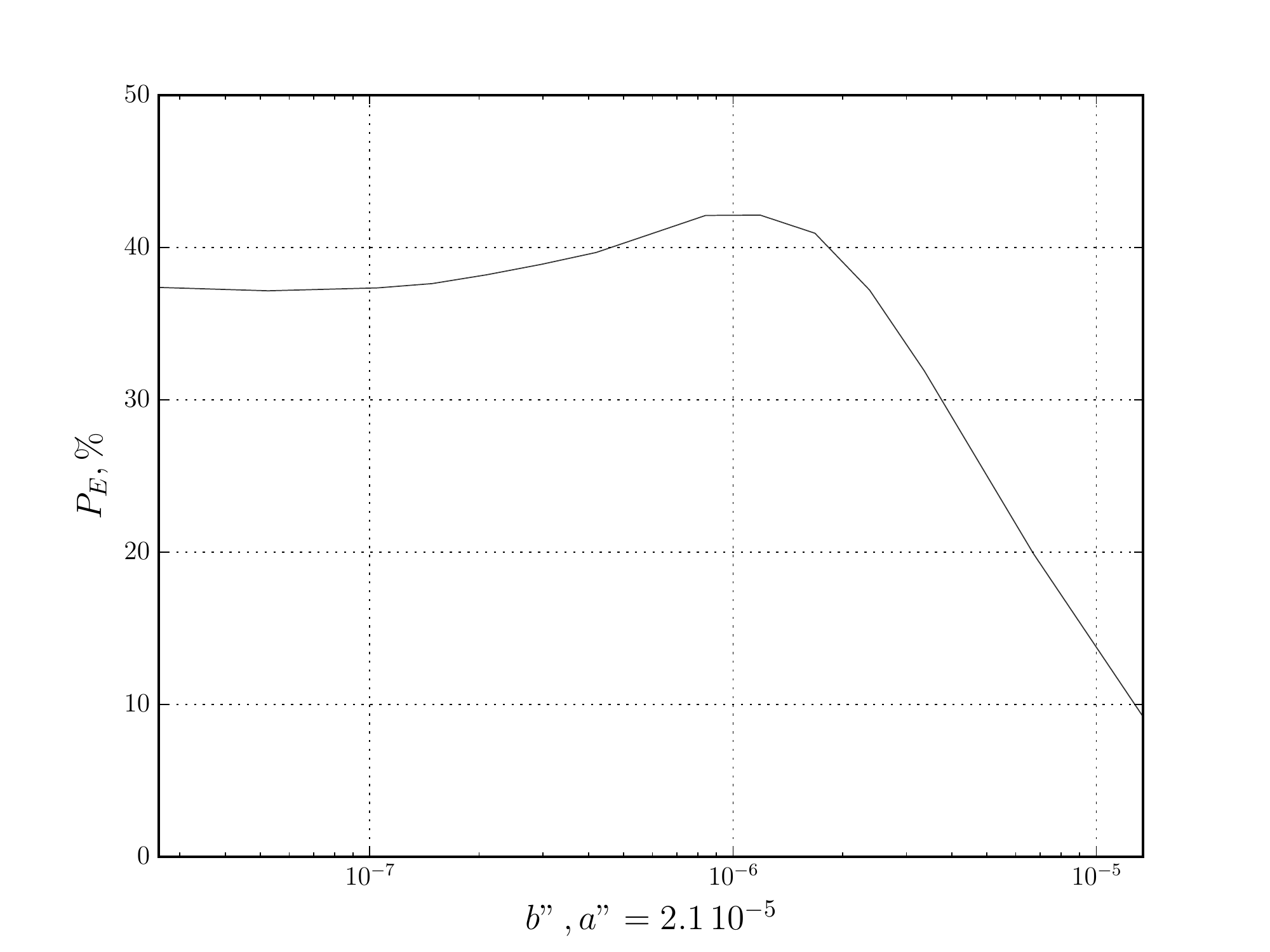}\tabularnewline
\end{tabular}
\par\end{centering}

\caption{Sensitivity w.r.t. the model parameters $a"$ and $b"$.\label{fig:Sensitivity-w.r.t.-the}}
\end{figure}

Figure~\ref{fig:Sensitivity-w.r.t.-the} depicts the sensitivity
of our scheme to the estimation of the sensor noise by computing the
classification error for different values of $a"$ and $b"$. We can
see that the estimation of the sensor noise is rather important, going
from example from $a"=2.1\,10^{-5}$ to $a"=10^{-5}$ increases the
detectability by approximately 5\%.

\subsection{Results with 8-bit embedding\label{sub:Problem-with-8}}

\begin{figure*}[t]
\begin{centering}
\begin{tabular}{ccc}
\includegraphics[width=0.3\textwidth]{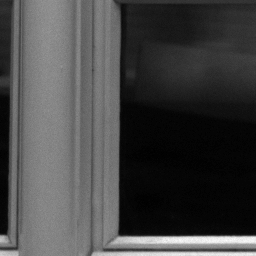} & \includegraphics[width=0.3\textwidth]{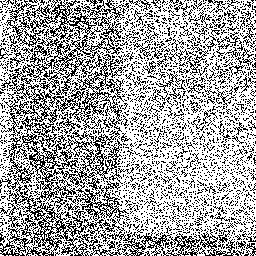} & \includegraphics[width=0.3\textwidth]{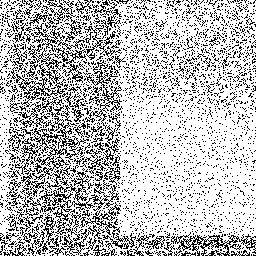}\tabularnewline
(a) & (b) & (c)\tabularnewline
\end{tabular}
\par\end{centering}

\caption{Portion of an image (a) and locations of embedding changes when 16-bit
to 8-bit is used (b) and when it is not used (c) (for better rendering,
inactivate interpolation on your pdf viewer).\label{fig:Portion-of-an}}
\end{figure*}

\begin{table}[h]
\begin{centering}
\begin{tabular}{|c|c|c|}
\hline 
 & NS & NS 8-bits\tabularnewline
\hline 
$P_{E}$ & \textbf{42.8\%} & 18.4\%\tabularnewline
\hline 
\end{tabular}
\par\end{centering}

\caption{Impact of using the side-information of NS.\label{tab:Results-8b}}
\end{table}
Table~\ref{tab:Results-8b} presents results when the embedding is
directly performed on 8-bit cover images and we can notice that in
this case the scheme becomes highly detectable. We explain this problem
by the fact that dark regions, which undergo both stego signal and
sensor-noise of small variance are not modified in a natural way.
These regions are especially sensitive to steganalysis because they
are less noisy than bright regions, and because the value of the photo-site
before 16-bit/8-bit conversion highly impact the sign of the embedding
change in this case. Figure~\ref{fig:Portion-of-an} show the embedding
change for a portion of a cover image having dark areas and we can
see that for the 8-bit embedding, the number of embedding changes
are less important since the dithering effect offered by the use of
the 16-bit image is lost here (the sensor noise is in this case centered
directly on the quantization cell). Trying to improve NS in this practical
setup is left for future researches.

\subsection{Gamma correction\label{sub:Gamma-correction-res}}

Table~\ref{tab:Perf-GAMMA} shows the detectability results of NS
once gamma correction is performed during the developing step both
on cover and on stego images. 16-bit cover images are used. Since
the model of the stego signal is adapted to feat the model of the
sensor-noise after the gamma correction (see~\ref{sub:Gamma-correction})
we can check that the undetectability of NS is still high. 

We see also that on MonoBase the embedding rate increases w.r.t. the
parameter $\gamma$. This is because for $\gamma>1$ the variance
of the stego signal increases for small photo-site values and decreases
for large photo-site values. The inverses occurs for $\gamma<1.$
In the first case this is due to the convexity of the transform, in
the second case to the concavity of the transform. On a database only
composed of bright images, the effect would be the opposite. 

\begin{table}[h]
\begin{centering}
\begin{tabular}{|c||c|c|c|c|c|}
\hline 
$\gamma$ & 2.5 & 2 & 1.5 & 1 & 0.5\tabularnewline
\hline 
\hline 
$P_{E}$ & 40.2\% & 42.1\% & 40.7\% & 42.2\% & 43.3\%\tabularnewline
\hline 
$E_{r}$(bpp) & 1.61 & 1.62 & 1.55 & 1.24 & 0.5\tabularnewline
\hline 
\end{tabular}
\par\end{centering}

\caption{Performances of NS after Gamma correction.\label{tab:Perf-GAMMA} }
\end{table}

\subsection{Downsampling\label{sub:Downsampling-and-scaling-res}}

Table~\ref{tab:Results-Scaling} presents the detectability results
after downsampling the images of a factor 2 (because the images were
already really small we did not use $c>2$). We can notice the detectability
is smaller after down-sampling, which is probably due to the effect
of the square root law~\cite{ker2008square}.

\begin{table}[h]
\begin{centering}
\begin{tabular}{|c|c|c|c|c|}
\hline 
 & \textbf{NS} & NS  & NS & NS\tabularnewline
 &  & Sub-sampling & $c=2$, Box & $c=2$, Tent\tabularnewline
\hline 
$P_{E}$ & \textbf{42.8\%} & 48\% & 47.7\% & 48.0\%\tabularnewline
\hline 
\end{tabular}
\par\end{centering}

\caption{Detectably after x2 downsampling.\label{tab:Results-Scaling}}
\end{table}

\begin{figure}
\begin{centering}
\includegraphics[width=1\columnwidth]{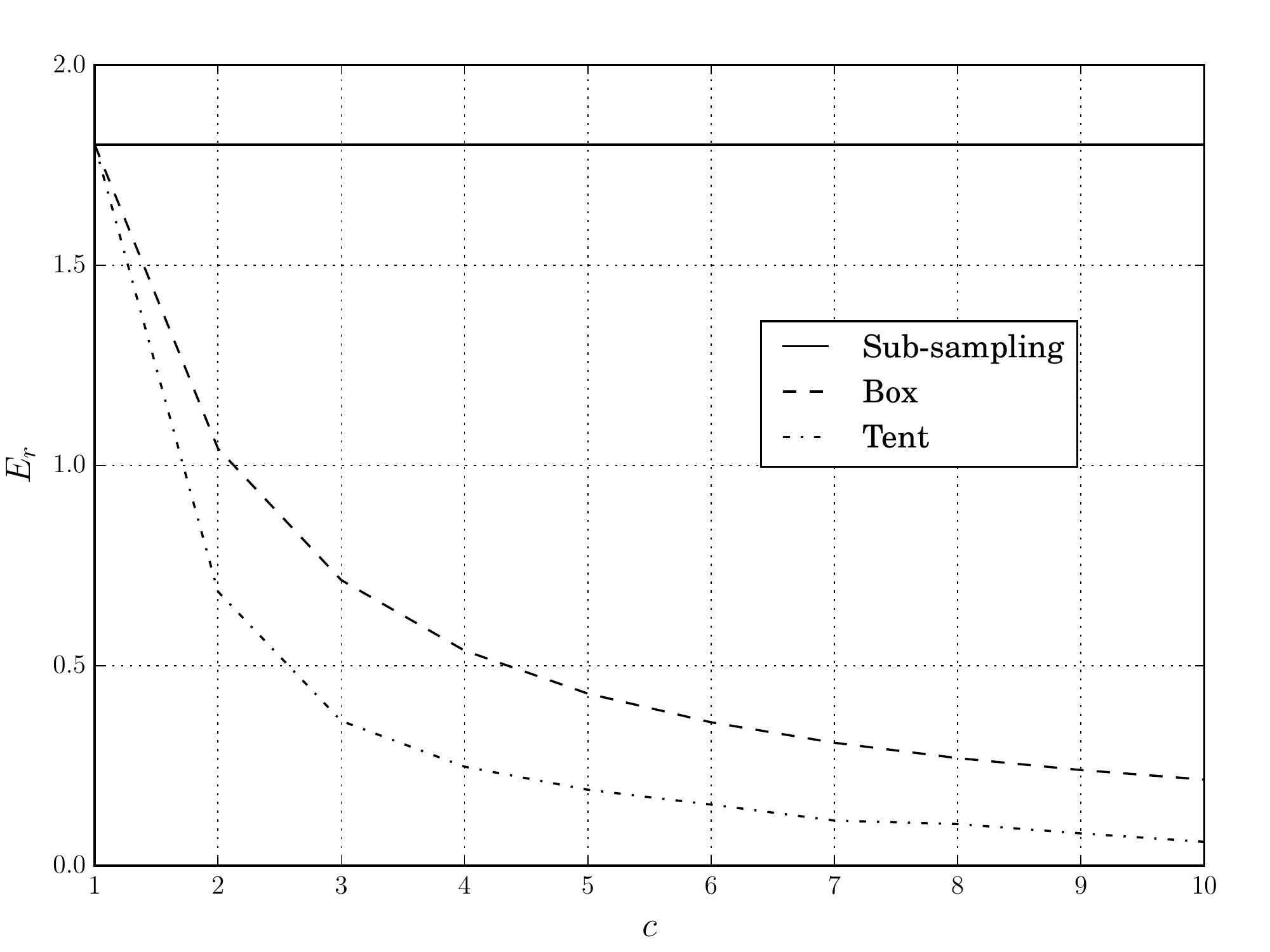}
\par\end{centering}

\caption{Embedding rate vs scaling factor, 1000 ISO toward 1250 ISO embedding
for a cover image uniformly distributed, $a'=2.1\times10^{-5}$,$b'=8.4\times10^{-7}$.\label{fig:Embedding-rate-vs}}
\end{figure}

Figure~\ref{fig:Embedding-rate-vs} presents the evolution of the
embedding rates computed using the densities of the stego signal for
the three down-sampling methods presented in (see~\ref{sub:Down-sampling-(and-up-sampling)})
for an image having photo-sites uniformly and independently distributed
between 0 and $2^{16}-1$. 

We can notice that the embedding rates rapidly decrease w.r.t. the
scaling factor for Box or Tent downscaling. The rate is constant for
classical sub-sampling but this method generates aliasing and is never
used in practice. If we for example look at the typical down-scaling
operation used in BOSS-Base, a 18MP image (3840 x 2592) was downsampled
with c=5, which lead in these case to $E_{r}\thickapprox0.4\,\mathrm{pbb}$
for Box downsampling and $E_{r}\thickapprox0.2\,\mathrm{pbb}$ for
Tent downsampling. Compared with the initial embedding rate of 1.8
pbb, the reduction is rather important. Note that for a given detectability
constraint, the embedder can always increase the values of $a"$ and
$b"$ to increase the payload, or change the cover-source switching
setup by using $ISO_{1}<1000$ or $ISO_{2}>1250$. 

We can draw two remarks comparing with the literature on steganography:
\begin{enumerate}
\item The evolution of the detectability is perfectly inline with the analyzed
effect of rescaling in steganalysis~\cite{kodovsky2013steganalysis}
(see for example Figure 1 of~\cite{kodovsky2013steganalysis}) which
outlines that the Tent kernel is more detectable than the Box kernel.
\item The generation of the stego signal after resampling implies computation
of conditional probabilities which force the embedding scheme to correlate
the neighboring embedding changes, which is also inline with the Synch
or CMD strategies presented in~\cite{denemark2014selection,li2015strategy}.
\end{enumerate}

\section{Another strategy: cover-source perturbation\label{sec:Another-strategy:-cover-source}}

We want to mention alternative strategy to cover-source switching
which is cover-source perturbation. In this case the embedding does
not mimic another cover-source but just slightly perturbs it. This
can simply done by setting $b"=0$, setting $a"$ and comparing cover
images with stego images with the same sensitivity (here 1000 ISO),
this way the stego signal slightly perturbs the sensor-noise. The
advantage of cover-source perturbation is the fact that is doesn't
require to model the source sensor noise which can be very interesting
in practice.

We present results related to cover-source perturbation in Table~\ref{tab:Res-Pertub}
which depicts the evolution of the detection error and the embedding
w.r.t. $a"$ when both cover images are at 1000 ISO. We can notice
that cover-source perturbation may offer undetectability but at the
price of a smaller embedding rate. For example for the same $P_{E}$
as NS with cover-source switching, the embedding rate is roughly divided
by 3. 

\begin{table}[h]
\begin{centering}
\begin{tabular}{|c||c|c|c|c|c|}
\hline 
a'' & $4\,10^{-7}$ & $1.5\,10^{-6}$ & $6.3\,10^{-6}$ & $2.5\,10^{-5}$ & $10^{-4}$\tabularnewline
\hline 
\hline 
$P_{E}$ & 46.2\% & 45.5\% & 41.4\% & 24.7\% & 5.4\%\tabularnewline
\hline 
$E_{r}$ (bpp) & 0.16 & 0.32 & 0.63 & 1.19 & 1.98\tabularnewline
\hline 
\end{tabular}
\par\end{centering}

\caption{Cover-source perturbation, comparison with covers at 1000 ISO. \label{tab:Res-Pertub}}
\end{table}

\section{Conclusions and perspectives}

We have proposed in this paper a new methodology for steganography
based on the principle of cover-source switching, i.e. the fact that
the embedding should mimics the switching from one cover-source to
another. The scheme we presented scheme (NS) used the sensor noise
to model one source and message embedding is performed by generating
a suited stego signal which enables the switch. This method, in order
to provide good undetectability performances while proposing high
embedding rates, has to use RAW images as inputs. We also show in
the paper how to handle different steps of image developing, including
quantization, gamma correction, color transforms and rescaling operations.

In future works we want also to investigate other setups for NS steganography,
such as choosing other ISO parameters and different camera models.
It will also be important to try to improve direct embedding on 8-bit
images and to address more practical implementation such as embedding
in the JPEG-domain. 

From the adversary point of view, we would like to see if more appropriate
feature could be designed for this category of schemes, this kind
of features should not be sensitive only to image variation, but also
to the sensor noise whose variance is function of the pixel luminance.

Another track of research is to consider other ways to perform cover-source
switching (or cover-source perturbation, see section~\ref{sec:Another-strategy:-cover-source}),
where the source can be represented here by, for example, the demosaicing
algorithm. Since the behavior of demosaicing algorithms fluctuates
a lot in textures, we think that this strategy would generate embedding
changes that are closer to the ones used currently by other steganographic
methods. 

Finally we hope that this methodology will page the road for new directions
in steganography.

\section{Acknowledgments}

The author would like to thank Boris Valet for his work on sensor
noise estimation, Cyrille Toulet and Matthieu Marquillie for their
help on the Univ-lille HPC, Remi Bardenet for his help on sampling
strategies, Tomas Pevny and Andrew Ker for their inspiring conversations
of the definition of the source, and CNRS for a supporting grant on
cyber-security.

\bibliographystyle{IEEEtran}


\end{document}